\documentclass[a4paper,12pt]{article}
\usepackage[utf8]{inputenc}
\usepackage{amsmath}
\usepackage{amssymb}
\usepackage{mathtools}
\usepackage{fancyhdr}
\usepackage{siunitx}
\usepackage{mleftright}
\usepackage{booktabs}
\usepackage{caption}
\usepackage{aligned-overset}
\usepackage{subcaption}
\usepackage[intlimits]{esint}
\usepackage{stackengine}
\usepackage[explicit]{titlesec}
\usepackage{tocloft}
\usepackage{lipsum}
\usepackage{authblk}
\usepackage[textwidth=15cm,textheight=22cm]{geometry}

\renewcommand{\headrulewidth}{0.4pt}

\DeclareMathAlphabet{\mathbbold}{U}{bbold}{m}{n}

\usepackage[hyphens,spaces,obeyspaces]{url}
\usepackage[pdftitle={A Note on and Generalization of Exploring Modified Kaniadakis Entropy: MOND Theory and the Bekenstein Bound Conjecture}, pdfauthor={Julius Lehmann},colorlinks=true, citecolor=blue, urlcolor=black, linkcolor=blue]{hyperref}
\usepackage[style=numeric]{biblatex}
\newcommand{\dr}[1]{\mathrm{#1}}

\newcommand{\der}[2][]{\frac{\mathrm{d}^{#1}}{\mathrm{d}#2^{#1}}}
\newcommand{\Der}[3][]{\frac{\mathrm{d^{#1}}#2}{\mathrm{d}#3^{#1}}}

\newcommand{\nexp}[1]{\dr{e}^{#1}}

\DeclarePairedDelimiter\klam{(}{)}

\DeclarePairedDelimiterX\braket[2]{\langle}{\rangle}{#1 \delimsize\vert #2}

\title{A Note on and Generalization of ``Exploring Modified Kaniadakis Entropy: MOND Theory and the Bekenstein Bound Conjecture''}
\author{Julius Lehmann\footnote{e-mail: \texttt{julius.lehmann(at)tum.de}}}
\affil{\textit{Physics of Complex Biosystems, Technical University of Munich, 85748 Garching, Germany}}
\date{\today}

\normalsize

\begin{document}

\maketitle

\begin{abstract}
   In a recent paper by Ambrósio et al.~\cite{ambrósio2024exploring}, it was shown that the gravitational force law of the Modified Newtonian Dynamics (MOND) phenomenology can be derived within the framework of entropic gravity and the holographic principle by assuming an entropy function other than the conventional Boltzmann-Gibbs entropy. In particular, they derived the standard interpolation function of MOND together with an analytical expression for the acceleration constant $a_0$ by utilizing Kaniadakis' modified entropy. In this short note, using the same methodology, we generalize this result and show that MONDian behavior is a rather general consequence of combining entropic gravity with non-Boltzmann-Gibbs entropies, which depends on only a few conditions imposed on the generalized entropy function.
\end{abstract}

\subsection{Introduction}

An outstanding problem in modern (astro)physics is the explanation of several gravitational observations that do not agree with our most successful description of gravity, Einstein's general theory of relativity (GR). Most prominently, the rotational speed $v$ of the majority of galaxies does not decrease with distance, but instead remains roughly at a limiting speed~\cite{Rubin1978,Bosma1981AJ} given by the empirical baryonic Tully-Fisher relation~\cite{Tully1977,Courteau1999ApJ}, $v^4\propto M_\dr{galaxy}$, where $M_\dr{galaxy}$ is the mass contained in the entire galaxy. To solve this ``flatness problem'' in galactic rotation curves, two competing models have been proposed in the literature: the first model relies on the introduction of undetected (and most likely non-baryonic) dark matter to account for the observations, while the second model modifies gravity.

A representative of the latter approach is Modified Newtonian Dynamics (MOND) \cite{Milgrom1,Milgrom2,Milgrom3} and was proposed in the 1980s by Mordehai Milgrom as alternative to dark matter. It suggests a modification of Newton's law of universal gravitation~\cite{footnoteMOND} in a regime of low accelerations, below a certain critical acceleration scale $a_0$. Introducing the Newtonian gravitational acceleration $a_\dr{N}\equiv\frac{GM}{R^2}$, where $G$ is the gravitational constant, $M$ is the central mass, and $R$ is the distance of a test particle from the central mass, we can write the modified force law as
\begin{equation}
    F_\dr{MOND}=m\mu\klam[\Big]{\frac{a_\dr{N}}{a_0}}a_\dr{N}.
\end{equation}
The function $\mu(x)$ interpolates between the standard Newtonian regime ($a_\dr{N}\gg a_0$) and the deep-MOND regime ($a_\dr{N}\ll a_0$). Consistency with Newtonian mechanics requires
\begin{align}
    \mu(x)&\to1,\quad x\gg1
    \intertext{while agreement with astronomical observations requires }
    \mu(x)&\approx x,\quad x\ll1.
\end{align}
This modification of the gravitational force law is able to fit the rotation curves of galaxies and explain the Tully-Fisher relation with remarkable accuracy. The modified force law also explains other astronomical observations besides galactic rotation curves; see Refs.~\cite{Famaey2011kh,Banik2021FromGB} for a comprehensive review of the observational phenomenology and predictions of MOND. Nevertheless, there are observational difficulties with MOND when confronted with clusters~\cite{2023A&Li,2003MNRASanders} or wide binaries~\cite{10.1093banik} (but see~\cite{Chae_2024} for contrary claims). The most important problem, however, is the lack of a fundamental theory that gives rise to MONDian gravity in the low-acceleration limit. While the main interest lies in a relativistic field theory analogous to GR with the correct deep-MOND regime, having a theoretical framework for non-relativistic, classical gravity could provide insight into the most promising path towards a covariant formulation of MONDian gravity.

In recent years, based on the discovery by Jacobson that the Einstein field equations can be viewed as a thermodynamic equation of state under a set of minimal assumptions~\cite{PhysRevLett.75.1260}, there has been great interest in theories that understand gravity as an entropic force, with particular impetus coming from the theory proposed by Verlinde in the 2010s~\cite{Verlinde2010hp}. There, and more generally in Refs.~\cite{Abreu_2017,modesto2010entropic}, the following expression was derived for the (entropic) gravitational force due to particles inside a region of space with a holographic screen and entropy $S$:
\begin{equation}\label{eq:GravF}
    F_\dr{G}=G\frac{Mm}{R^2}\frac{4\ell_\dr{P}^2}{k_\dr{B}}\Der{S}{A}.
\end{equation}
$A$ is the area of the holographic screen, assumed to be a sphere, $A=4\pi R^2$, $\ell_\dr{P}$ is the Planck length, and $k_\dr{B}$ is Boltzmann's constant. Using the well-known expression for the black hole entropy, $S_\dr{BH}=k_\dr{B}\frac{\pi R^2}{\ell_\dr{P}^2}$~\cite{1975CMaHawking}, recovers the usual expression for the gravitational force, $F_\dr{G}=\frac{GMm}{R^2}$. Already in~\cite{Abreu_2017} and recently in~\cite{ambrósio2024exploring}, it was shown that by basing the entropy of the holographic screen not on the conventional Boltzmann-Gibbs entropy,
\begin{equation}
    S_\dr{BG}=-k_\dr{B}\sum_ip_i\ln(p_i),
\end{equation}
where $p_i$ is the probability of being in state $i$, but instead on generalized entropies like the Tsallis entropy~\cite{1988JSPTsallis},
\begin{equation}
    S_\dr{T}=-k_\dr{B}\frac{1-\sum_ip_i^q}{1-q},
\end{equation}
or the Kaniadakis entropy~\cite{PhysRevE.66.056125},
\begin{equation}\label{eq:KaniadakisS}
    S_\dr{K}=-k_\dr{B}\sum_ip_i\frac{p_i^{q-1}-p_i^{1-q}}{2(1-q)},
\end{equation}
MONDian corrections to the Newtonian gravitational effects emerge. Note that $0<q<1$ can be seen as a measure of additivity of the system, and that in both cases we recover the usual Boltzmann-Gibbs entropy in the limit $q\to1$.

Let us briefly sketch the derivation by Ambrósio et al.~\cite{ambrósio2024exploring} to get an overview of the framework used. Starting from Eq.~\eqref{eq:KaniadakisS} and using the microcanonical ensemble, where all
states have equal probability $p_i=\frac{1}{\Omega}$ with $\Omega$ the total number of microstates, the expression simplifies to
\begin{equation}
    S_\dr{K}=k_\dr{B}\frac{\Omega^{1-q}-\Omega^{q-1}}{2(1-q)}.
\end{equation}
Setting this equal to the black hole entropy, they find for the number of microstates:
\begin{equation}
    \Omega=\klam[\bigg]{\!(1-q)\frac{S_\dr{BH}}{k_\dr{B}}+\sqrt{1+(1-q)^2\klam[\Big]{\frac{S_\dr{BH}}{k_\dr{B}}}^2}}^{\frac{1}{1-q}}.
\end{equation}
Substituting this expression for $\Omega$ into the Boltzmann-Gibbs entropy gives a modified version for the Kaniadakis entropy:
\begin{equation}
    S_\dr{K}^\ast=\frac{k_\dr{B}}{1-q}\ln\klam[\bigg]{\!(1-q)\frac{S_\dr{BH}}{k_\dr{B}}+\sqrt{1+(1-q)^2\klam[\Big]{\frac{S_\dr{BH}}{k_\dr{B}}}^2}}.
\end{equation}
Using this entropy together with Eq.~\eqref{eq:GravF}, they arrive after some algebra at the expression
\begin{equation}
    F_\dr{MOND}=\frac{ma_\dr{N}}{\sqrt{1+(\frac{a_\dr{N}}{a_0})^2}},
\end{equation}
with the Newtonian gravitational acceleration as defined at the beginning and the critical acceleration scale
\begin{equation}
    a_0=(1-q)\frac{\pi GM}{\ell_\dr{P}^2}.
\end{equation}
This expression for the force clearly satisfies the conditions of MOND, and the resulting interpolation is known as the ``standard'' interpolation function.

In the next section, we will demonstrate how this ansatz can be used to compute the generalized entropy of the holographic screen based on black hole thermodynamics. In particular, we will give conditions for the generalized entropy to recover MONDian behavior.

\subsection{Generalized Entropy and MOND}
There exist numerous non-extensive generalizations of the standard Boltzmann-Gibbs statistics. Two of these were previously mentioned and are referred to as Tsallis statistics and Kaniadakis statistics. All generalized statistics have in common that they recover the conventional statistics in some limit. We define the generalized $q$-entropy as
\begin{equation}\label{eq:GenS}
    S\{L_q\}=-k_\dr{B}\sum_iL_q(p_i),
\end{equation}
depending on an unspecified function of probability $L_q(x)$. The function must be chosen such that we recover the Boltzmann-Gibbs entropy in the $q\to1$ limit:
\begin{equation}
    \lim_{q\to1}S\{L_q\}=S_\dr{BG}=-k_\dr{B}\sum_ip_i\ln(p_i),
\end{equation}
which implies the relation
\begin{equation}
    \lim_{q\to1}L_q(x)=x\ln(x).
\end{equation}
Furthermore, to behave as entropy, we require $L_q(0)=L_q(1)=0$, i.\,e. inaccessible states do not contribute to $q$-entropy and a singular state does not possess $q$-entropy. Another useful property of the $q$-entropy is subadditivity of two systems:
\begin{equation}
    \sum_{i,j}L_q(p_1(i)p_2(j))\leq\sum_{i,j}L_q(p_1(i))+\sum_{i,j}L_q(p_2(j)),
\end{equation}
which implies that $L_q(x)$ is a convex function.

With this definition of the generalized entropy, we will follow the approach of~\cite{ambrósio2024exploring} and consider first the microcanonical ensemble. Taking $p_i=\frac{1}{\Omega}$ we get the expression
\begin{equation}
    S\{L_q\}=-k_\dr{B} L_q\klam[\Big]{\frac{1}{\Omega}}\Omega,
\end{equation}
where we used $\sum_i1=\Omega$. The next step is to equate the $q$-entropy to the black hole entropy and then use the resulting number of microstates for the Boltzmann-Gibbs entropy. This is equivalent to finding the modified entropy $S^\ast$ that satisfies the implicit equation
\begin{equation}\label{eq:DefEq}
    -k_\dr{B} L_q\klam[\Big]{\nexp{-\frac{S^\ast}{k_\dr{B}}}}\nexp{\frac{S^\ast}{k_\dr{B}}}=S_\dr{BH}.
\end{equation}
To compute the effect of this entropy on the gravitational force, we are mainly interested in the derivative of $S^\ast$ with respect to the area. Implicitly differentiating the expression gives
\begin{equation}\label{eq:DerS}
    \Der{S^\ast}{A}=\frac{k_\dr{B}}{4\ell_\dr{P}^2}\frac{1}{L_q'\klam[\big]{\nexp{-\frac{S^\ast}{k_\dr{B}}}}+\frac{S_\dr{BH}}{k_\dr{B}}},
\end{equation}
where we write $L_q'\klam[\big]{\nexp{-\frac{S^\ast}{k_\dr{B}}}}\equiv\der{x}L_q(x)\Bigr|_{x=\nexp{-\frac{S^\ast}{k_\dr{B}}}}$. Substituting this expression into the expression for the entropic force~\eqref{eq:GravF}, we further find
\begin{equation}
    F_\dr{MOND}=\frac{ma_\dr{N}}{L_q'\klam[\big]{\nexp{-\frac{S^\ast}{k_\dr{B}}}}+\frac{S_\dr{BH}}{k_\dr{B}}}=m\mu\klam[\Big]{\frac{a_\dr{N}}{a_0}}a_\dr{N}.
\end{equation}
Since we are not interested in the exact form of the MONDian force, but only in the conditions on $L_q(x)$ such that we recover the standard Newtonian law on the one hand and the deep-MOND behavior on the other hand, we restrict our analysis to the two limit cases $a_\dr{N}\to0$ (deep-MOND) and $a_\dr{N}\to\infty$ (Newton). To compute the limits, note that $S^\ast$ is a function of the black hole entropy $S_\dr{BH}$, which in turn can be written as function of Newtonian gravitational acceleration:
\begin{equation}
    S_\dr{BH}=k_\dr{B}\frac{\pi GM}{a_\dr{N}\ell_\dr{P}^2}.
\end{equation}
With this, we can identify the limits as $a_\dr{N}\to0\implies S_\dr{BH}\to\infty$ and $a_\dr{N}\to\infty\implies S_\dr{BH}\to0$. First, consider the Newtonian limit. There, we can compute the value of $S^\ast$ from the implicit equation~\eqref{eq:DefEq} and find
\begin{equation}
    \nexp{-\frac{S^\ast(0)}{k_\dr{B}}}=1
\end{equation}
which implies $S^\ast(0)=0$. Combining this with the condition $\mu\klam{\frac{a_\dr{N}}{a_0}}\to1$, we get the condition on $L_q(x)$ of:
\begin{equation}
    L_q'(1)=1.
\end{equation}
To compute the deep-MOND limit, note that $S^\ast$ and $S_\dr{BG}$ should behave similarly, since both functions are convex and approach each other in the limit $q\to1$. Therefore, it is reasonable to assume that $S_\dr{BH}\to\infty$ also implies $S^\ast\to\infty$. This in turn implies $\nexp{-\frac{S^\ast}{k_\dr{B}}}\to0$, so we can approximate $L_q(x)$ for small values of $x$. For that purpose, assume $L_q(x)\sim C_qx^{\alpha_q}\ (x\to0)$ for some constants depending on $q$, then it follows:
\begin{equation}
    S^\ast\approx\frac{k_\dr{B}}{1-\alpha_q}\ln\klam[\Big]{-\frac{S_\dr{BH}}{C_qk_\dr{B}}}.
\end{equation}
Inserting this into the denominator of the ``entropic'' interpolation function, we find
\begin{equation}
    L_q'\klam[\Big]{\nexp{-\frac{S^\ast}{k_\dr{B}}}}+\frac{S_\dr{BH}}{k_\dr{B}}\approx(1-\alpha_q)\frac{S_\dr{BH}}{k_\dr{B}},
\end{equation}
which in turn gives the correct limiting behavior whenever $\alpha_q<1$:
\begin{equation}
    \mu\klam[\Big]{\frac{a_\dr{N}}{a_0}}=\frac{1}{L_q'\klam[\big]{\nexp{-\frac{S^\ast}{k_\dr{B}}}}+\frac{S_\dr{BH}}{k_\dr{B}}}\approx \frac{a_\dr{N}\ell_\dr{P}^2}{(1-\alpha_q)\pi GM}=\frac{a_\dr{N}}{a_0},
\end{equation}
upon identifying the acceleration constant
\begin{equation}
    a_0=(1-\alpha_q)\frac{\pi GM}{\ell_\dr{P}^2}.
\end{equation}
Similarly, if we assume $L_q(x)\sim C_qx^{\alpha_q}\ln(x)\ (x\to0)$, we find the same asymptotic expressions for $\mu(x)$ and $a_0$. Given that all $L_q(x)$ are assumed to be convex, the interpolation function defined by it will transition monotonously between these two regimes. This, together with the behavior in both the Newtonian and the deep-MOND limit, fulfills the tenets of modified Newtonian dynamics. An immediate consequence of our findings is the non-existence of MONDian behavior for the case of standard Boltzmann-Gibbs entropy: there we have $L_q(x)=x\ln(x)$, implying for the exponent $\alpha_q=1$, and the expression diverges.

From these sets of conditions we have that the more popular candidates for generalized entropies, the Tsallis and Kaniadakis entropies, exhibit MONDian behavior in the low-acceleration limit (in the employed framework). We find the following functional forms of $L_q(x)$ for the respective entropies:
\begin{align}
    L_q^\dr{T}(x)&=x\frac{x^{q-1}-1}{q-1}\intertext{and}
    L_q^\dr{K}(x)&=x\frac{x^{q-1}-x^{1-q}}{2(q-1)}.
\end{align}
In both cases we have agreement with the conditions presented in the previous paragraph, which is consistent with the result obtained in Refs.~\cite{Abreu_2017,ambrósio2024exploring}.

\subsection{Conclusion}
In this note, using the same methodology, we generalized a result of Ambrósio et al. presented in~\cite{ambrósio2024exploring}, which demonstrated the emergence of MONDian behavior in the entropic gravity framework by modifying the underlying entropy function. Starting from a generic form of entropy, $S\{L_q\}=-k_\dr{B}\sum_iL_q(p_i)$, we derived the conditions necessary for MONDian behavior to appear and showed that this is a rather general consequence of combining entropic gravity with non-Boltzmann-Gibbs entropies. In particular, the set of conditions (and assumptions) on $L_q(x)$ are: (1) $L_q(x)$ is convex, (2) $L_q(0)=L_q(1)=0$, (3) $L_q'(1)=1$, and (4) $L_q(x)\sim x^{\alpha_q}$ or $L_q(x)\sim x^{\alpha_q}\ln(x)$ for $0\leq\alpha_q<1$ when $x\to0$. It is a simple task to check that two of the more widely used generalized entropies, the Tsallis and Kaniadakis entropies, satisfy these conditions.

In the present paper, we have limited ourselves to the general consequences that follow from using the framework as given in~\cite{ambrósio2024exploring}. Whether entropic forces are a correct description of gravity, whether MOND is the correct low-acceleration limit of gravity in general, or whether some other variation of this framework can account for this was beyond the scope of this paper. Nevertheless, the observation that MONDian effects in gravity are hard to avoid in the employed framework of entropic gravity whenever the underlying statistics of the holographic screen are not of Boltzmann-Gibbs type might give way to a better description of gravity and a covariant relativistic theory of modified Newtonian dynamics.

\clearpage

\lhead{}
\renewcommand{\headrulewidth}{0.0pt}
\printbibliography

\end{document}